\documentclass[aps,prl,twocolumn,10pt,showpacs,superscriptaddress,amsmath,amssymb]{revtex4-1}
\usepackage{graphicx}
\usepackage{amsmath}  % needed for \tfrac, \bmatrix, etc.
\usepackage{amsfonts} % needed for bold Greek, Fraktur, and blackboard bold
\usepackage{dcolumn}% Align table columns on decimal point
\usepackage{bm}% bold math
\usepackage{color,xspace}

\newcommand{\slfcr}{sc}
\newcommand{\xsc}[1]{x^{\text{(sc)}}_{#1}}

\begin{document}

\title{Creating Entanglement Using Integrals of Motion}

\author{Maxim Olshanii}
\affiliation{Department of Physics, University of Massachusetts Boston, Boston, MA 02125, USA}
\email{Maxim.Olchanyi@umb.edu}

\author{Thibault Scoquart}
\affiliation{D\'{e}partement de Physique, Ecole Normale Sup\'{e}rieure, 24, rue Lhomond, 75005 Paris, France}
\affiliation{Department of Physics, University of Massachusetts Boston, Boston, MA 02125, USA}

\author{Dmitry Yampolsky}
\affiliation{Department of Physics, University of Massachusetts Boston, Boston, MA 02125, USA}

\author{Vanja Dunjko}
\affiliation{Department of Physics, University of Massachusetts Boston, Boston, MA 02125, USA}
\email{Vanja.Dunjko@umb.edu}

\author{Steven Glenn Jackson}
\affiliation{Department of Mathematics, University of Massachusetts Boston, Boston Massachusetts 02125, USA}

\date{\today}

\begin{abstract}
A quantum Galilean cannon is a 1D sequence of $N$ hard-core particles with special mass ratios, and a hard wall; conservation laws due to the reflection group $A_{N}$ prevent both classical stochastization and quantum diffraction. It is realizable through specie-alternating mutually repulsive bosonic soliton trains. We show that an initial disentangled state can evolve into one where the heavy and light particles are entangled, and propose a sensor, containing $N_{\text{total}}$ atoms, with a $\sqrt{N_{\text{total}}}$ times higher sensitivity than in a one-atom sensor with $N_{\text{total}}$ repetitions.
%02.30.Jr			Partial differential equations
%03.65.Fd		Algebraic methods (see also 02.20.-a Group theory)
%02.30.Ik			Integrable systems
%03.65.Ca	Formalism
%32.80.Qk 	Coherent control of atomic interactions with photons
%02.30.Jr   Partial differential equations
%67.85.-d 	Ultracold gases, trapped gases
%67.85.De	Dynamic properties of condensates; excitations, and superfluid flow
%03.65.Fd 	Algebraic methods (see also 02.20.-a Group theory)
%05.45.Mt	Quantum chaos; semiclassical methods
%06.20.Dk	Measurement and error theory
%
\end{abstract}

\maketitle

% =======================================================================
% Introduction
% =======================================================================
In a 1D system of hard-core particles, by tuning the ratios between the particle masses one can choose a variety of distinct regimes of motion \cite{hwang2015_467,redner2004_1492}.
While generic values result in thermalization, for
some special ones the system maps to known multidimensional
kaleidoscopes \cite{gaudin1983_book,olshanii2015_105005}. The outcome velocities then become determined by the initial velocities only, independent of the initial positions if the particle ordering is preserved.
In the quantum version, the eigenstates are finite superpositions of plane waves, with no diffraction
\cite{sutherland2004_book}. In particular a (run in reverse) Galilean cannon---a hard wall followed by a 1D sequence of $N$ hard-core particles with mass ratios of
$1:\frac{1}{3}:\frac{1}{6}:\frac{1}{10}:\ldots:\frac{2}{N(N+1)}$---will evolve
from a state where only the lightest particle is moving (approaching the others from infinity), to a state where
all of the particles are moving away from the wall,
with the same speeds. The final state of the heavy particles is very different from the initial
one, yet highly predictable, presenting an opportunity: if created, a superposition of the final and initial states would be a Schr\"{o}dinger cat state \cite{haroche_book}: just as the $\alpha$-particle controls the well-being of the cat, so the state of the lightest particle controls either (a) the rest of the particles or (b) the heaviest particle, after the others are detected. And yet the whole system is in a pure state.  The stored entanglement stored is ready to be used, and we will propose an interferometric application.

% =======================================================================
% General solution
% =======================================================================

%Consider a system of $N$ 1D hard-core particles with masses $m_{i}$ ($i=1,\,\ldots,\,N$), $x_{i}>0$, and a hard wall at $x=0$. The Hamiltonian is given by the kinetic energy,

Consider a system of $N$ 1D hard-core particles with masses
$m_{1},\,m_{2},\,\ldots,\,m_{N}$ on the half-line $x>0$, bounded by a hard wall at $x=0$. The Hamiltonian is given by the kinetic energy,
\begin{align}
\hat{H} = -\sum_{i=1}^{N} \frac{\hbar^2}{2 m_{i}} \frac{\partial^2}{\partial x_{i}^2}
\,\,;
\label{H}
\end{align}
the wavefunction $\Psi(x_{1},\,x_{2},\,\ldots,\,x_{N})$ satisfies the boundary conditions
\begin{align}
\Psi\Big|_{x_{1}=0} = \Psi\Big|_{x_{1}=x_{2}}=\ldots=\Psi\Big|_{x_{N-1}=x_{N}}=0
\,\,.
\label{bc}
\end{align}
The coordinate transformation $x_{i}=\sqrt{\mu/m_{i}}\,z_{i}$ for $i=1,\,2,\,\ldots,\,N$, where $\mu$ is
an arbitrary mass scale that can be chosen at will, converts the system
to a single $N$-dimensional particle of mass $\mu$ moving inside a mirror-walled wedge formed by
$N$ mirrors; the outward normalized normals to its mirrors are given by $\bm{n}_{1}=-\bm{e}_{1}$
and $\bm{n}_{i}=\sqrt{m_{i}/(m_{i-1}+m_{i})} \bm{e}_{i-1} - \sqrt{m_{i-1}/(m_{i-1}+m_{i})} \bm{e}_{i}$
for $i=2,\,3,\,\ldots,\,N$, where $\bm{e}_{i}$ is the unit vector along the $z_{i}$-axis.

For a generic set of masses, sequential reflections about the mirrors generate an infinite set of spatial transformations. However, for every full reflection group \footnote{\ldots{a}s distinct from a subgroup thereof}  of a regular multidimensional polyhedron
(i.e.\ Platonic solid), there is a set of masses whose corresponding system of mirrors generates that group
 \cite{olshanii2015_105005}. In these
cases, the eigenstates of the system can be found exactly through Bethe ansatz \cite{gaudin1983_book,olshanii2015_105005}: they are given by finite linear combinations of
plane waves.
In particular, the ``Galilean cannon''
set of masses $m_{1},\,m_{2}=m_{1}/3,\,m_{3}=m_{1}/6,\,\ldots,\,m_{N}=\frac{2}{N(N+1)} m_{1}$
corresponds to the symmetry group
of a regular $N$-dimensional tetrahedron \footnote{Another class of systems identified in Ref.~\cite{olshanii2015_105005}, not considered here, corresponds
to the cases where the system is bounded by two hard walls---a finite box.}.

\begin{figure}%[h!]
\centering
\includegraphics[scale=.65,clip=true]{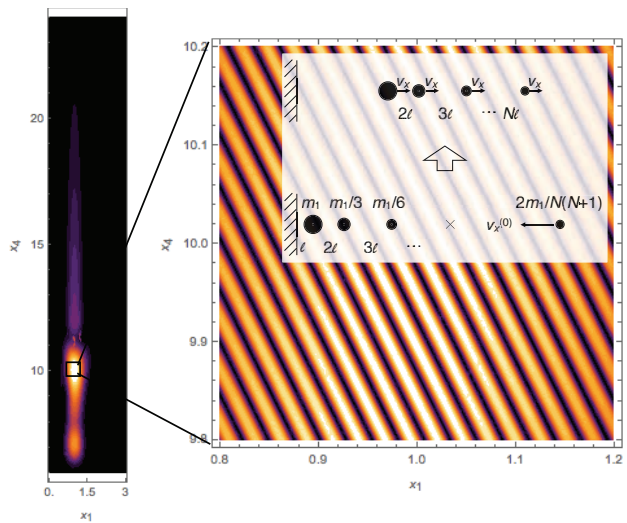}
\caption{
A Schr\"{o}dinger cat state produced in a Galilean cannon system with $N=4$ particles.
The conditional two-body
density distribution for the heaviest and the lightest particles, $|\Psi(x_{1},\,3\ell,\,6\ell,\,x_{4})|^2$, is plotted, at a time
$t_{\text{obs}}$, subject to the
second and the third particles being held at $(x_{2},\, x_{3})=(3,\,6)\,\ell$. The initial state is a
Gaussian wavepacket for each of the particles. The mean initial velocities of all the particles but the lightest
vanish, while the lightest is approaching the system at a speed $\mathcal{V}_{x}^{(0)}$. The magnified
portions are centered around the self-crossing point,
$(x_{1},\, x_{2},\,x_{3},\, x_{4})=(1,\,3,\,6,\,10)\,\ell$, where the head of the distribution
is capable of crossing its tail. In the figures, the head part is not visible outside of the
crossing with the tail.
The centers of the initial Gaussians are at
$(x_{1}^{(0)},\,x_{2}^{(0)},\,x_{3}^{(0)},\,x_{4}^{(0)}) = (1,\,3,\,6,\,31.5)\,\ell$.
The speed $\mathcal{V}_{x}^{(0)}$, the observation time, and
the dispersions of the initial Gaussians  are
$\mathcal{V}_{x}^{(0)} = 949\, \hbar/(m_{1}\ell)$, $t_{\text{obs}} = 0.0332\,m_{1}\ell^2/\hbar$,
and
$(\sigma_{x,1},\,\sigma_{x,2},\,\sigma_{x,3},\,\sigma_{x,4}) = (0.129,\,0.223,\,0.315,\,9.49)\,\ell$
respectively. With the exception of the lightest particle direction, the initial spatial distribution is round, if
expressed through the $z$-coordinates; it is disproportionately elongated in the $z_{4}$ direction (i.e. the initial state of the lightest particle is comparatively much broader) to ensure that
the many-body wavepacket can self-cross.
The observation time corresponds to the exact middle point of the time evolution
(see main text). \textbf{Inset:} Galilean cannon run in reverse: the lightest particle sets in motion all
the heavy particles, at the same speed.
In addition, the initial positions of the
particles are chosen in such a way that there exists a particle configuration---namely the
initial positions of all particles but the lightest, whose position is instead indicated by the cross---that is realized \emph{twice} in the course of the time evolution.
             }
\label{f:density_BB}
\end{figure}

As the initial state for the reverse Galilean cannon (Fig.~\ref{f:density_BB}, insert),
we take the tensor product of the Gaussian wavepackets for each particle,
supplemented by its images needed to satisfy the boundary conditions (\ref{bc}).
The corresponding solution of the time-dependent
Schr\"{o}dinger equation with the Hamiltonian (\ref{H}), subject to the boundary conditions (\ref{bc}) is
\begin{align}
\begin{split}
&
\Psi(z_{1},\,z_{2},\,\ldots,\,z_{N})
=
\\
&
\quad
\left(
\sum_{\hat{g}} (-1)^{\mathcal{P}(\hat{g})} \hat{g}
\right)
\prod_{i=1}^{N} \psi(z_{i},\, t\,|\,z_{i}^{(0)},\,v_{z,i},\,\sigma_{z,i})
\,\,,
\end{split}
\label{Psi}
\end{align}
where
\begin{multline*}
\psi(z,\, t\,|\,z^{(0)},\,v_{z},\,\sigma_{z})
\\
=  A(t)
e^
{
-
\frac
  {
   (z-z^{(0)})^2 - 4 i  \hbar \mu \sigma_{z}^2 v_{z} (z-z^{(0)}) + 2 i \mu t v_{z}^2 \sigma_{z}^2/\hbar
  }
  {
   4 \sigma_{z}^2 [1 + (i \hbar t)/(2 \mu  \sigma_{z}^2) ]
  }
  }
\end{multline*}
is the one-body Gaussian wavepacket expanding freely;
$A(t) = \left(2 \pi \sigma_{z}^{2} [1 + (i \hbar t)/(2 \mu  \sigma_{z}^2)]^{2}\right)^{-1/4}$ is the normalization constant;
the operators $\hat{g}$
are the transformations of space generated by the sequential application of reflections about
the $N$ mirrors given by the normals $\bm{n}_{i}$ described above. This solution is obtained in the same way as the Bethe eigenstates in the case of hard-wall kaleidoscopes
\cite{olshanii2015_105005}, which, in turn, is a generalization of a general solution
for kaleidoscopes with Robin's boundary conditions \cite{gaudin1983_book,gutkin1979_6057,sutherland1980_1770,gutkin1982_1,emsiz2006_191,emsiz2009_571,emsiz2010_61}, which was inspired by the Bethe ansatz solutions for a gas of bosons  \cite{girardeau1960_516,lieb1963_1605,mcguire1963_622,gaudin1971_386}. The set
of $\hat{g}$'s forms the reflection group $A_{N}$, containing $(N+1)!$ elements \cite{humphreys_book_1997}.
$\mathcal{P}(\hat{g})$ is the parity of the group element, i.e.\ the parity of the number of
reflections about the $N$ generating mirrors (given in particular by the normals $\bm{n}_{i}$
above) needed to produce this element.

% =======================================================================
% Cat
% =======================================================================

Having obtained the general solution for the problem, let us return to the physical
coordinates $x_{1},\, x_{2},\,\ldots\,\,x_{N}$. For the sequence depicted in the inset of Fig.~\ref{f:density_BB} (``the Galilean cannon run in reverse''),
the initial velocities of the particles are  $v_{x,N} = -\mathcal{V}_{x}^{(0)} < 0$ and $v_{x,1}=v_{x,2}=\ldots=v_{x,N-1}=0$.
At the final stage of the process, each particle moves away from the wall with the same speed $\mathcal{V}_{x} = \sqrt{m_{N}/M} \mathcal{V}_{x}^{(0)} = \mathcal{V}_{x}^{(0)}/N$,
where $M=\sum_{i=1}^{N} m_{i}$ is the total mass of the system. The initial distances between the particles,
and between the leftmost particle and the wall, are assumed much greater than the widths of their initial packets:
$x_{1}^{(0)} \gg \sigma_{x,1}$ and $x_{i}^{(0)} - x_{i-1}^{(0)} \gg \mbox{max}(\sigma_{x,i},\,\sigma_{x,i-1})$ for
$i=2,\,3,\,\ldots,\,N$. In this case, the initial state is close
to a \emph{product} state of individual non-overlapping states of finite support: the images,
while formally present in the expression (\ref{Psi}), will be exponentially small at $t=0$ (but
will come to prominence at later times, as the particle wavepackets move around and broaden).

If the initial multidimensional Gaussian wavepacket is sufficiently long, the various parts of the superposition (\ref{Psi}) will start overlapping at
intermediate stages of the time evolution,
forcing the particles to entangle---despite the closeness of the initial state to a product state.
The most promising is the superposition between 1. the initial packet and 2. the ``outgoing'' one, where all the particles are moving with a velocity $+\mathcal{V}_{x}$: here the state of the lightest particle
controls the state of each of the heavier ones, including the heaviest. It turns out that for a properly tuned set of the initial conditions,
there will be regions of space where these two waves \emph{spatially overlap} and all other parts of (\ref{Psi}) are exponentially small.
Indeed, one can show that if a classical trajectory passes through the point
$
(\xsc{1},\,\xsc{2},\,\ldots,\,\xsc{N})
=
(\ell,\, 3\ell,\,6\ell,\,\ldots,\,\frac{N(N+1)}{2}\ell)
$, it will do so \emph{twice}, once during the initial leg of the evolution and once during the final.
(Here and below, $\ell$ is an arbitrary length scale, and the subscript '$\text{\slfcr}$' stands for ``self-crossing.'')
The distances between the particles increase linearly with the index: $\xsc{j}-\xsc{j-1}=j\ell$.
%$
%(\xsc{1}-0,\,\xsc{2}-\xsc{1},\,
%\ldots,\,\xsc{N}-\xsc{N-1})
%\stackrel{\mu=m_{1}}{=}
%(\ell,\, 2\ell,\,\ldots,\,N\ell)
%$.
At the exact middle point of the time evolution (which can be shown to equal the time when the lightest particle would hit the wall if there were
no other particles present), the state around the point of self-crossing is close to
\mbox{}
\vspace{-\baselineskip}
\mbox{}
\begin{minipage}[t]{0.5\textwidth}
\begin{multline*}
\Psi(x_{1},\,x_{2},\,\ldots,\,x_{N})
\propto
e^{-i(m_{N} \mathcal{V}_{x}^{(0)} (x_{N}-\xsc{N} ) + \frac{\phi}{2})}
\\+e^{+i(\sum_{i=1}^{N-1} m_{i} \mathcal{V}_{x} (x_{i}-\xsc{i}) + m_{N} \mathcal{V}_{x} (x_{N}-\xsc{N} )+\frac{\phi}{2})}\,,
\end{multline*}
\mbox{}
\vspace{-\baselineskip}
\mbox{}
\end{minipage}
where the relative phase $\phi$ can be approximated, using the eikonal approximation, as $\hbar\phi= 2 m_{1}  \mathcal{V}_{x}^{(0)}  \ell$. In the state above, the coordinate of the lightest particle is entangled with the center-of-mass position for the
remaining bodies. Below, we will use this entanglement as a way to improve the sensitivity of interferometric measurements.
As for the ``cat'' per se, we have the position of the center of mass of the particles being spread over an $\sim N^2 \ell$ range. If this seems too abstract, one can also generate entanglement between two localized objects, one light and one heavy: suppose
the intermediate particles $2,\,3,\,\ldots,\,N-1$ have been detected at particular positions. The particles 1 and $N$ remain entangled, in spite of the \mbox{$N-2$} hard walls between them formed by the detected intermediate particles. For example,
when the intermediate particles are detected at their ``self-crossing'' values, the state of the system becomes
\begin{multline*}
%&
\Psi_{1,\,N}(x_{1},\,x_{N})
\propto
e^{-i(m_{N} \mathcal{V}_{x}^{(0)} (x_{N}-\xsc{N} ) + \frac{\phi}{2})}
\\
+e^{+i(m_{1} \mathcal{V}_{x} (x_{1}-\xsc{1}) + m_{N} \mathcal{V}_{x} (x_{N}-\xsc{N} )+\frac{\phi}{2})}
\,\,.
\end{multline*}
This state is a paradigmatic Schr\"{o}dinger cat state where a light particle ($x_{N}$), the  ``$\alpha$-particle'', is entangled
with a heavy one ($x_{1}$), the ``cat''.

Figure~\ref{f:density_BB} shows the results of time propagation according to the above scheme, for $N=4$.
%
%{\color{blue}
At the classical self-crossing point, the incident and the outgoing waves dominate:
the clear interference fringes in
the $x_{1}$--$x_{4}$ plane are a signature of that. The absence of diffraction is a sign of integrability. Also note that for a generic set of masses,
the most probable outcome of the process is the equipartition of energy. In this case, the velocity of the heaviest particle will be $\sqrt{N(N+1)/2} \stackrel{N \gg 1}{\sim} N$ times lower than in the integrable case, leaving it effectively at rest.

We have numerically computed
the R\'{e}nyi entropy $S_{2}[\mbox{particle 1}] \equiv -\ln[\mbox{Tr}[\hat{\rho}_{\text{particle 1}}^2] ]$ for the reduced density
matrix $\hat{\rho}_{\text{particle 1}}$ of the heaviest particle, for the state truncated to the square
area in Fig.~\ref{f:density_BB}, and $(x_{2},\, x_{3})$ fixed to
$(3,\,6)\,\ell$. To compute the entropy, we discretized the $(x_{1},\,x_{4})$ space into a square
grid, and in doing so, reduced the computation to a standard setting where the
Hilbert space has a finite number of dimensions.
We verified that for small enough grid spacing, the entropy does not depend on
the value of the spacing. As expected, $S_{2}[\mbox{particle 1}]  \approx \ln(1.991)$, close to $\ln 2$,  indicating two
element-wise-distinct sets of particle momenta.
%}

% =======================================================================
% 4D fringes
% =======================================================================

\begin{figure}%[h!]
\centering
\includegraphics[scale=.45,clip=true]{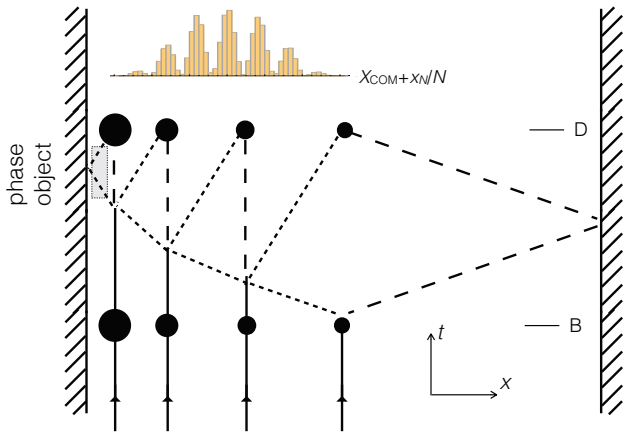}
\caption{
A Galilean-cannon-based interferometer: an example with $N=4$ particles.
Initially all particles are at the point of self-crossing,
$(x_{1},\, x_{2},\,x_{3},\, x_{4})=(1,\,3,\,6,\,\ldots,\,\frac{N(N+1)}{2} \stackrel{N=4}{=}10)\,\ell$.
At time B, a beam-splitter is applied to the lightest particle. After a series of collisions, each particle
returns to its initial position, at the detection time D, where the particle positions are measured.
A phase object, both spatially and temporarily localized, is shown as a grey rectangle.
While no individual particle position distribution possesses any structure, the
distribution of the $\tilde{X} \equiv X_{\text{COM}} + \frac{1}{N} x_{N}$
variable shows interference fringes, whose position is controlled by the phase introduced by the object.
The inset shows a sample histogram derived from 79900 simulated realizations of the detection cycle:
those were selected from a longer sequence;
only the realizations with particle positions
within a hypercube of dimensions $0.1\ell \times 0.1\ell \times 0.1\ell \times 0.1\ell$ centered
at the self-crossing point $(1,\,3,\,6,\,10)\,\ell$ were kept. The distance between the crests is consistent
with the predicted value of $\Delta \tilde{X}  =  2\pi/(M \mathcal{V}_{x}) = 0.0165\,\ell$. The
velocity kick induced by the beam-splitter is the same as the initial velocity $\mathcal{V}_{x}^{(0)}$
of  Fig.~\ref{f:density_BB}. The initial wavepacket width of the lightest particle is
$\sigma_{x,4} = 0.408\,\ell$. The period of time between the beamslitting and detection is
$0.0210\,m_{1}\ell^2/\hbar$. The remaining parameters are the same
as in Fig.~\ref{f:density_BB}. Note that thanks to the presence of the
beamsplitter and an additional mirror, the initial wavepacket is no longer required to be long enough to self-cross.
Accordingly, if viewed in the $z$-coordinates, the initial many-body wavepacket is perfectly round.
}
\label{f:4D_fringes_BB}
\end{figure}

Notice that, if plotted as a
function of the coordinate $\tilde{X} \equiv X_{\text{COM}} + \frac{1}{N} x_{N}$, the $N$-body density
corresponding to the state $\Psi(x_{1},\,x_{2},\,\ldots,\,x_{N})$ in the vicinity of the self-crossing point
shows interference fringes with crest-to-crest distance of $\Delta \tilde{X}  =  2\pi/(M \mathcal{V}_{x})$,
as if produced by a wave for a single massive particle of mass $M$ split by a $\pm M \mathcal{V}_{x}/2$ beamsplitter.
Here, $X_{\text{COM}} \equiv \sum_{i=1}^{N} m_{i} x_{i}/M$ is the center of mass coordinate.
Figure~\ref{f:4D_fringes_BB} shows a sample interferometric scheme exploiting this effect.
In this scheme, the \emph{beamsplitter only acts on the lightest particle}, while due to the entanglement buildup,
the more massive particles also affect the position of the fringes.

\newcommand{\Nlightest}{\mathcal{N}_{\text{light}}\xspace}
\newcommand{\Nheaviest}{\mathcal{N}_{\text{heavy}}\xspace}
\newcommand{\Ntotal}{\mathcal{N}_{\text{total}}\xspace}
%{\color{blue}
Below we will suggest a possible experimental realization in which the role of the particles is played by polymers of atoms---bosonic solitons. Assume that the lightest particle is a polymer consisting of $\Nlightest$ atoms.
The heaviest one will contain
$ \Nheaviest = \Nlightest \times N(N+1)/2   \sim \Ntotal$ atoms, where
$\Ntotal$
is the total number of atoms in the system. Imagine that
a phase object, a potential barrier of height $U$, \emph{per atom}, acting during a limited time $\tau$, is introduced
between the wall and the default position of the heavy polymer. The best sensitivity to the strength  $U$
of the phase object can be easily estimated as
$\Delta U_{\text{Galilean cannon}} \sim \hbar/\left(\tau  \Nheaviest\right) \sim \hbar/\left(\tau \Ntotal\right)$,
i.e.\ by the energy-time uncertainty relation for  an interferometer formed by the heaviest polymer
attempting to measure a barrier of hight $\Ntotal U$ within a time $\tau$.
However, if $\Ntotal$ individual atoms  are used, the maximal sensitivity is only
$\Delta U_{\text{individual atoms}} \sim \hbar/\left(\tau \sqrt{\Ntotal}\right)$, which is
the sensitivity of a single-atom interferometer further improved by the signal-to-noise reduction using
$\Ntotal$ repetitive measurements.
The net relative sensitivity
gain produced by the entanglement, for a given number of atoms $\Ntotal$ available,
becomes a $\textit{gain} \sim \sqrt{\Ntotal}$.
%}
%
%

% =======================================================================
% Conclusion and Experimental
% =======================================================================

In conclusion, we showed that for a particular one-dimensional mass sequence, it is possible to
realize a protocol in which the system evolves, on its own, from a product state to a state where
a heavy particle becomes entangled with a light one, thus realizing Schr\"{o}dinger's
a-cat-and-an-$\alpha$-particle paradigm. We show numerically that the R\'{e}nyi entropy of the
heavy particle can rise to almost $\ln 2$. The robustness  of the protocol
is due to the integrability of the model that protects it from both classical stochastization and quantum
diffraction. We suggest a concrete way to exploit the heavy-light entanglement by proposing an
atomic interferometric sensor scheme that shows an $\sqrt{\Ntotal}$ increase in sensitivity, where $\Ntotal$ is the total number of atoms employed.

As an empirical realization of the scheme presented above we suggest using chains of cold bosonic
solitons \cite{khaykovich2002_1290,strecker2002_150,cornish2006_170401}.
For our scheme, it is necessary to have two  internal states available (or, alternatively, two kinds of atoms).
We assume that like species attract each other, while the scattering length between the opposite
species is tuned to a positive value. For $\mbox{Li}^{7}$ atoms, the desired window in Feshbach magnetic field strength does exist: in particular,
at $855\, \mbox{G}$, the scattering lengths governing a $(m_{F}=-1)$--$(m_{F}=0)$ mixture are
$a_{-1,-1} \approx -0.5\, a_{\text{B}}$, $a_{0,0} \lessapprox -10\, a_{\text{B}}$,
and $a_{-1,0} \approx +1.0\, a_{\text{B}}$, where $a_{\text{B}}$ is the Bohr radius \cite{Randy_Lithium}.
One would have to ensure that the kinetic energy of the relative motion of the solitons must be lower than both the intra- and inter-specie interaction energy
per particle, to  ensure both a suppression of the inelastic effects and an absence of inter-specie transmission.
Finally, the soliton sizes must be adjusted to fit the desired mass sequence.
Nontrivial integrals of motion present, at the mean-field level, in cold one-dimensional Bose gases
may provide a way to accurately divide
the gas onto desired fractions \cite{zakharov1972_62,satsuma1974_284,dunjko2015_150100075},
exact in the mean-field limit. For instance,
the $1,\,\frac{1}{3},\,\frac{1}{6},\,\frac{1}{10}$ mass spectrum considered above can be created using three types of quench of the coupling constant:
sudden increase by a factor of $4$, $\frac{25}{4}$, and $9$. Starting from a single soliton of a mass ${\cal M}$, a sequence of quenches
$9 \to \frac{25}{4} \to \frac{25}{4} \to 4 \to 4 \to 4$ would lead to an ensemble of eight solitons of masses
$\left\{\frac{2}{75},\frac{2}{45},\frac{2}{25},\frac{4}{45},\frac{1}{9},\frac{2}{15},\frac{16}{75},\frac{4}{15}\right\} \times {\cal M}$ of the mass of the
original soliton, with $\frac{8}{225}  \times {\cal M}$ lost to the thermal atoms.
If the third, fifth, sixth, and seventh members of this sequence are further removed, the resulting group
$\left\{\frac{2}{75},\frac{2}{45},\frac{4}{45},\frac{4}{15}\right\} \times {\cal M}$ constitutes the desired sequence.

\begin{figure}%[h!]
\centering
\includegraphics[scale=.55,clip=true]{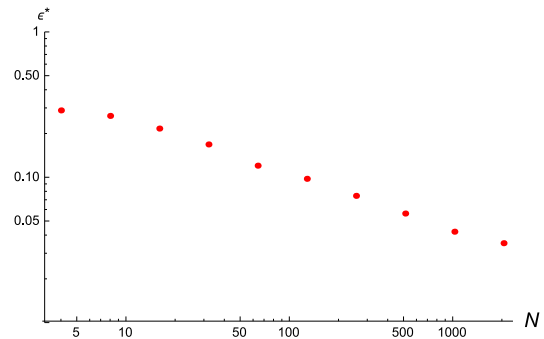}
\caption{
Relative size of the mass fluctuation of the lightest particle at which a transition from an integrable to stochastic behavior occurs,
as a function of the number of particles.
             }
\label{f:epsStar_vs_N}
\end{figure}

Potentially, residual fluctuations in the soliton occupations may be detrimental to the effects we discuss. To address this problem,
we performed a series of classical simulations of the dynamics of a Galilean cannon with masses fluctuating from one run to another.
For a given particle, the variance of its mass fluctuations was proportional to the mean particle mass, mimicking the Poissonian  law,
while the mass distribution itself had a rectangular profile. The spectra of the mean masses was the same as the mass spectra considered above.
The parameter $\epsilon \equiv \text{StDev}[m_{N}]/\text{Mean}[m_{N}]$ controlled the overall magnitude of the mass fluctuations. In the integrable
limit, $\epsilon \to 0$, the final velocities of the particles are equal. On the other hand, for finite values of $\epsilon$, one expects to see, on average, an equipartition
of energy. As a quantitative definition of the critical value $\epsilon^{\star}$ that signifies the transition between the integrable and stochastic regimes,
we choose the value of $\epsilon$ at which 
\newcommand{\energ}{E\xspace}
\newcommand{\veloc}{v\xspace}
\newcommand{\infTAve}[1]{\langle #1 \rangle_{\!\infty}}
$S^{\energ}=S^{\veloc}$, where: $S^{\energ}$ is the spectral entropy of the heaviest-lightest pair,
$S^{\energ}\equiv -\sum_{n=1,\,N} q^{\energ}_{n} \ln q^{\energ}_{n}$, with $q^{\energ}_{n} = 
%\left(\text{Mean}\left[E_{n}\right]/\sum_{n'=1,\,N} %\text{Mean}\left[E_{n'}\right]\right)_{t\to\infty}$, 
\infTAve{E_{n}}/\sum_{n'=1,\,N} \infTAve{E_{n'}}$ and
$E_{n}=m_{n} v_{n}^2/2$, where $\infTAve{\cdots}$ is the infinite time average (this entropy is expected to be maximized whenever a system is stochastic); and $S^{\veloc}$ is the ``velocity'' entropy of the pair,
$S^{\veloc}\equiv -\sum_{n=1,\,N} q^{\veloc}_{n} \ln q^{\veloc}_{n}$, with $q^{\veloc}_{n} =
%\left(\text{Mean}[v_{n}]/\sum_{n'=1,\,N} %\text{Mean}[v_{n'}]\right)_{t\to\infty}$, 
\infTAve{v_{n}}/\sum_{n'=1,\,N} \infTAve{v_{n'}}$
(in our case, this entropy is maximized in the integrable regime, because then all final velocities are equal). Our results (see Fig.~\ref{f:epsStar_vs_N})
show that up to 1000 particles, mass fluctuations less than 5\% can be tolerated.
%

%%%%%%%%%%%%%%%%%%%%%%%%%%%%%%%%%%%%%%%%%%%%%%%%%%%%%%%%%%%%%%
% Acknowledgement

The authors thank Randy Hulet, H\'{e}l\`{e}ne Perrin, and Christopher Fuchs for help and comments.
This work was supported by the US National Science Foundation Grant No.\ PHY-1402249,
the Office of Naval Research Grant N00014-12-1-0400, and
a grant from the {\it Institut Francilien de Recherche sur les Atomes Froids} (IFRAF).
Financial support for TS provided by the Ecole Normale Sup\'{e}rieure
is also appreciated.

%%%%%%%%%%%%%%%%%%%%%%%%%%%%%%%%%%%%%%%%%%%%%%%%%%%%%%%%%%%%%%

%\bibliographystyle{spphys}       % APS-like style for physics

%\bibliography{Bethe_ansatz_v012,thermalization_literature_027,galilean_cannon_paper,Nonlinear_PDEs_and_SUSY_028}

%

\end{document}